\documentclass{jpp}
\usepackage{graphicx}
\usepackage{epstopdf, epsfig}
\usepackage{natbib}
\usepackage{amssymb}
\usepackage{amsgen}
\usepackage{amsfonts}
\usepackage{amsbsy}
\usepackage{amsmath}

\shorttitle{Dust strings in the PK-4 experiment}
\shortauthor{L. S. Matthews, et. al.}

\title{Effect of ionization waves on dust chain formation in a DC discharge}

\author{L.~S.~Matthews\aff{1} \corresp{\email{lorin\_matthews@baylor.edu}},
        K.~Vermillion\aff{1},
        P.~Hartmann\aff{1,2},
        M.~Rosenberg\aff{3},
        S.~Rostami\aff{1},
        E.~G.~Kostadinova\aff{1,4},
        T.~W.~Hyde\aff{1},
        M.~Y.~Pustylnik\aff{5},
        A.~M.~Lipaev\aff{6,7}, 
        A.~D.~Usachev\aff{6}, 
        A.~V.~Zobnin\aff{6}, 
        M.~H.~Thoma\aff{8}, 
        O.~Petrov\aff{1,6,7}, 
        H.~M.~Thomas\aff{5}
        \and O.~V.~Novitskii\aff{9}}

\affiliation{\aff{1}CASPER, Baylor University, One Bear Place 97316, Waco, TX 76798-7316, USA
\aff{2}Institute for Solid State Physics and Optics, Wigner Research Centre for Physics, P.O.Box. 49, H-1525 Budapest, Hungary
\aff{3}Department of Electrical and Computer Engineering, University of California San Diego, La Jolla, California 92093, USA
\aff{4}Physics Department, Auburn University, Auburn, Alabama, 36849, USA
\aff{5}Institut f{\"u}r Materialphysik im Weltraum, Deutsches Zentrum f{\"u}r Luft- und Raumfahrt (DLR), M{\"u}nchener Stra{\ss}e 20, 82234 We{\ss}ling, Germany
\aff{6}Institute for High Temperatures, Russian Academy of Sciences, Izhorskaya 13/19, 125412 Moscow, Russia
\aff{7}Moscow Institute of Physics and Technology, Institutsky Lane 9, Dolgoprudny, Moscow Region, 141700 Russia
\aff{8} I. Physikalisches Institut, Justus-Liebig-Universit{\"a}t Gie{\ss}en, Heinrich-Buff-Ring 16, 35392 Gie{\ss}en, Germany
\aff{9} Gagarin Research and Test Cosmonaut Training Center, 141160 Star City, Moscow Region, Russia
}

\begin{document}

\maketitle

\begin{abstract}
An interesting aspect of complex plasma is its ability to self-organize into a variety of structural configurations and undergo transitions between these states. A striking phenomenon is the isotropic-to-string transition observed in electrorheological complex plasma under the influence of a symmetric ion wakefield.  Such transitions have been investigated using the Plasma Kristall-4 (PK-4) microgravity laboratory on the International Space Station (ISS).  Recent experiments and numerical simulations have shown that, under PK-4 relevant discharge conditions, the seemingly homogeneous DC discharge column is highly inhomogeneous, with large axial electric field oscillations associated with ionization waves occurring on microsecond time scales. A multi-scale numerical model of the dust-plasma interactions is employed to investigate the role of the electric field on the charge of individual dust grains, the ion wakefield, and the order of string-like structures. Results are compared to dust strings formed in similar conditions in the PK-4 experiment.
\end{abstract}

\section{Introduction}

The PK-4 system is the latest generation in the line of microgravity dusty plasma experiments currently in operation on-board the International Space Station (ISS). Since its installation in the Columbus module in 2014, the PK-4 experiment has produced remarkable experimental data related to dust particle enhanced plasma emission \citep{Usachev16,Usachev18}, transverse ionization instability \citep{Zobnin16}, transformations of dust structures \citep{Polyakov17}, electrorheological and demixing phenomena \citep{Dietz17}, particle kinetics \citep{Liu18}, structural phase transitions \citep{Dietz18}, and dust density waves \citep{Jaiswal18}. Detailed reviews of past and recent microgravity dusty plasma activities can be found in \cite{Dietz18,Thomas19}.

\begin{figure}
\begin{center}
\includegraphics[width=0.6\columnwidth]{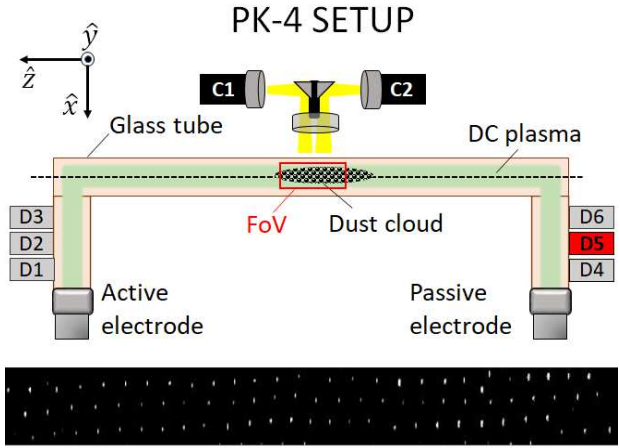}
\caption{(top) Schematic of the PK-4 experiment. Six microparticle dispensers (D1-D6) are mounted on the sides. Cameras C1 and C2 each have a field of view of 22.4 $\times$ 16.8 mm$^2$ and can be moved along as well as across the plasma chamber axis. (bottom) Dust particles within the PK-4 experiment showing the formation of chains.}
\label{fig:PK4setup}
\end{center}
\end{figure}

Besides these fundamental physical investigations, analysis of the raw experimental data has shown that under some circumstances the dust particles show a tendency for chain formation where the particles align into lines several tens of particles long parallel to the discharge tube axis, as reported in \cite{Pustylnik16,Schwabe19} and shown in figure~\ref{fig:PK4setup}. This happens most often (but not exclusively) when the polarity switching is applied, in which the  positive and negative polarities of the DC electrodes are alternated at a frequency of typically 100-500 Hz, with the aim of stabilizing the dust cloud in the field of view of the observing cameras. 

Several previous experiments have produced structures with aligned grains. Dust lane formation has been reported, e.g., during phase separation in binary complex plasmas under microgravity \citep{Sutterlin09,Du12}, driven by the electrostatic interaction between the charged dust grains in relative motion. Vertical dust particle chains can routinely be prepared in the electrode sheath region of a radio frequency (RF) gas discharge \citep{Kong11,Kong14,Mudi16}, where particle alignment is stabilized by the enhanced horizontal confinement provided by an open glass box and the ion wake field due to the fast (supersonic) streaming of ions around the particles~\citep{Hutchinson11,Hutchinson12,Kompaneets16}. The electrorheological effect (or the homogeneous-to-string transition) can also favor dust chain formation as demonstrated by \cite{Ivlev08,Ivlev11}. In this case the dust particles are surrounded by the quasi-neutral plasma bulk, but due to an externally applied alternating electric field and consequently streaming (subsonic) ions, the Debye screening sphere around the dust particles becomes distorted leading to an anisotropic inter-particle repulsion. Note that this is different than the electrorheological effect in granular suspensions, which results from polarization of the grains themselves \citep{Kwon15}.

Among these known chain-forming processes, the electrorheological effect is the most probable one to be acting in the positive column region of the PK-4 discharge plasma. For a PK-4 neon discharge at $p = 50$~Pa and $I = 1$~mA, the experimentally determined plasma parameters yield an axial electric field  $E_z \simeq 2.2\pm 0.2$~V/cm,  with an electron density $n_{\rm e} \simeq (2.2\pm 0.2)\times 10^8$~cm$^{-3}$ and electron temperature $T_{\rm e} \simeq 7 \pm 0.5 $~eV \citep{Usachev04,Khrapak12}. Assuming a stable positive column and based on the well-studied equilibrium transport behavior of Ne$^+$ ions in neutral Ne gas \citep{Skullerud90}, one can estimate the ion drift velocity to be $v_{\rm id} \simeq 190$~m/s resulting in a thermal Mach-number $M_{\rm th} = v_{\rm id}/v_{\rm th} = 0.54$. Here the ion thermal velocity is defined as $v_{\rm th} = \sqrt{k_{\rm B}T_{\rm i}/m_{\rm i}}$ assuming a temperature of $T_{\rm i} = 300$~K for the neon ions. The thermal Mach number is the key quantity for the estimation of the strength of the electrorheological effect based on the formula derived in \cite{Ivlev08} for the pairwise interparticle interaction energy
\begin{equation}
\label{IvlevPotential}
    W(r,\theta) \simeq \frac{Q^2}{4\pi\varepsilon_0}\left[ \frac{{\rm e}^{-r/\lambda_{\rm D}}}{r} - 0.43\frac{M_{\rm th}^2\lambda_{\rm D}^2}{r^3}\left( 3 \cos^2\theta -1 \right) \right],
\end{equation}
where $r$ is the distance between two dust grains of charge $Q$ aligned in the direction of the ion flow,  $\theta$ is the angle relative to the ion drift direction and $\lambda_{\rm D}$ is the unperturbed Debye screening length. In this description the isotropic Yukawa (screened Coulomb) interaction is modified by a dipole-like term and higher order contributions are neglected. It has been shown in \cite{Ivlev08} that anisotropy in the particle distribution gradually starts to develop above a critical value of the thermal Mach-number $M_{\rm cr} \simeq 0.3$ depending on the plasma conditions and that apparent ordered chains build up at $M_{\rm th}>1$ with increasing length and stability as the ion drift speed is further increased. Based on these previous findings and the assumption of a stable DC positive column, it could be expected that given the typical PK-4 conditions discussed above, the estimated thermal Mach number of 0.54 is insufficient to result in the formation of long particle chains, in contrast with the observed particle behavior.  However, recent simulations and experiments have shown that the plasma column supports fast-moving ionization waves, with associated ion flows speeds $M_{th} > 1$. Although these variations in the plasma occur on the micro-second timescale, they appear to have an influence on the dynamics of the dust grains, which typically occur on a millisecond timescale. 

In this work, we examine conditions affecting dust chain structure formation in the PK-4 experiment based on realistic gas discharge modeling, dust particle charging simulations, and calculations of the dust-dust and dust-ion interactions. Of particular interest is the strong electric field created by ionization waves which travel through the discharge column with a period on a microsecond timescale. A description of the PK-4 experiment and plasma conditions determined by a numerical model of the gas discharge are given in Section~2, with a description of the molecular dynamics (MD) model of the ion and dust dynamics in Section~3.  The dust charge and configuration resulting from applying different time-averaged discharge conditions are given in Section~4.  These results are compared with observations from the PK-4 experiment in Section~5.  Concluding remarks are given in Section~6.

\section{Methods}

The PK-4 experiment utilizes a long direct current (DC) discharge with an active length of approximately 400~mm in a glass tube with inner diameter of 30~mm, equipped with both neon and argon gases \citep{Pustylnik16}. The experiment utilizes several tools for manipulation of the dust, including movable radio frequency coils, a heating ring (thermal manipulator), an auxiliary internal ring electrode (electrical manipulation), and a 20~W continuous infrared laser (optical manipulation), which makes the system very versatile. The DC drive is realized with a high voltage arbitrary waveform generator with a frequency bandwidth up to 3~kHz, needed for applying polarity switching to the electrodes. Six dust particle dispensers are available, each filled with different mono-disperse spherical dust grains made of melamine-formaldehyde (MF). In the experiment, the dust particles are suspended in the center region of the discharge tube, in the bulk of the positive column. The observation of the dust ensemble and discharge glow is realized by video imaging, using a set of CCD cameras with an image resolution of 14.2 $\mu$m per pixel ~\citep{Schwabe19}. A detailed description of the setup and early experiments can be found in \cite{Pustylnik16}.

\subsection{Gas discharge modeling}

A cylindrical symmetric 2D Particle in cell with Monte Carlo collisions (PIC/MCC) code was implemented and used to simulate the motion and collisions of electrons and Ne$^+$ ions in neon gas and at solid surfaces in a DC discharge matching the PK-4 operating conditions. The electric field within the discharge tube is determined self-consistently from the boundary conditions at the electrodes and walls of the glass cylinder and the densities of the charged species.  The simulation was used to determine the plasma characteristics within the PK-4 experiment for a DC plasma in neon held at a pressure of $p = 40$~Pa, gas temperature $T_g = 300$~K, discharge current $I= 0.8$~mA (with approximately 1000~V DC voltage) with optional 500~Hz polarity switching. A detailed description of the model, its implementation and experimental verification are presented in a separate publication \citep{Hartmann_pk4}.

\begin{figure}
\begin{center}
\includegraphics[width=0.6\columnwidth]{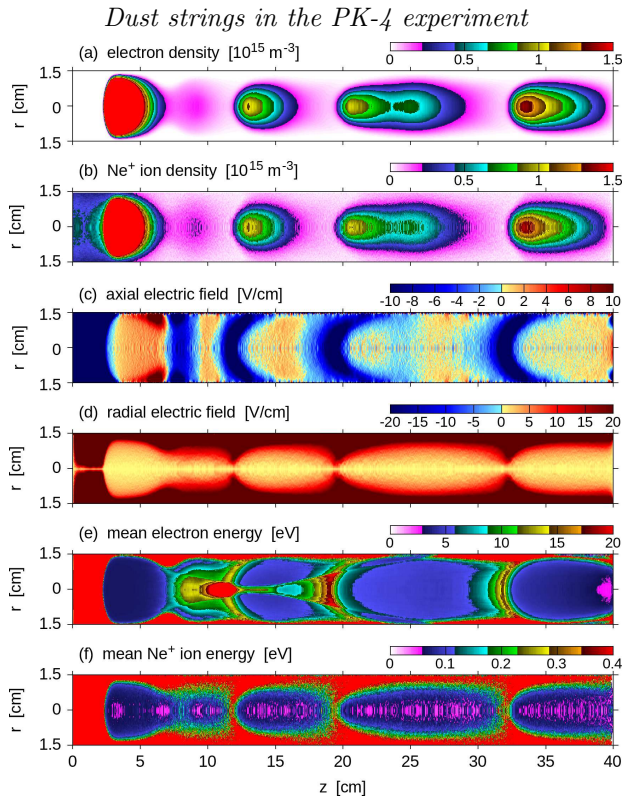}
\caption{Computed spatial distributions of plasma parameters: electron density (a), Ne$^+$ ion density (b), axial electric field (where positive indicates in the direction of increasing $z$) (c), radial electric field (d), mean electron energy (e), mean Ne$^+$ ion energy (f) at $p=40$\,Pa and $I=0.8$\,mA with the cathode at $z=0$. The data acquisition time was set to a very short $0.25\,\mu$s. The real aspect ratio of 3:40 is scaled by stretching the radial axis by a factor of two for better visibility.}
\label{fig:PIC1}
\end{center}
\end{figure}

Figure~\ref{fig:PIC1} shows the instantaneous spatial distribution of selected plasma parameters. The global structure reproduces the traditional structure of long DC discharges: a short cathode fall with large electric field, followed by a low field region with even a reversed field feature, and the small field positive column down to the anode. A dominant feature of the instant global structure is the presence of ionization waves which develop on a $\mu$s-time scale and travel along the column with phase velocities ranging between 500~m\,s$^{-1}$ and 1200~m\,s$^{-1}$. These quasiperiodic waves are characterized by a large amplitude modulation of the charged particle densities figure~\ref{fig:PIC1}(a,b) and alternating axial electric fields figure \ref{fig:PIC1}(c). A detailed analysis of the global plasma parameters computed with the same simulation under similar discharge conditions is presented in \cite{Hartmann_pk4}. The time-averaged  plasma parameters in the central region are $n_e$ = $n_i$ = $2.1 \times 10^{14}$~m$^{-3}$, mean energies $\langle\epsilon\rangle_e = 4.4$~eV and $\langle\epsilon\rangle_i= 0.04$~eV, and electric field $E = 245$~V/m. The presence of high amplitude ionization waves along the positive column makes the time-dependence of the plasma parameters at a given position (where the dust grains reside) of interest.

Here we focus on the local plasma environment in the central region of the discharge at position $z = 200$~mm and $r = 0$.  In the following graphs the time-dependence of the plasma parameters is shown with 0.25~$\mu$s resolution covering 250 $\mu$s total time at the central position of the cylinder. As shown in figure \ref{fig:PIC1} (a), the axial electric field varies in magnitude having a small positive value between the ionization waves (about 100~V/m, where positive indicates in the direction of increasing z) and peaking at about -2000~V/m as an ionization front passes.  

\begin{figure}
\begin{center}
\includegraphics[width=0.5\columnwidth]{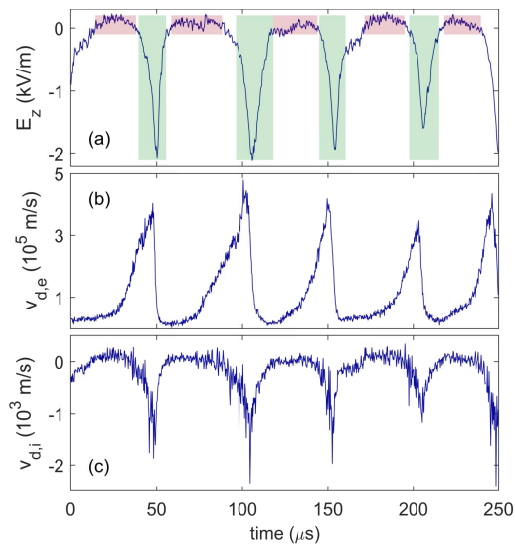}
\caption{(a) Axial electric field at the center of the column. Drift velocity of (b) electrons and (c) ions. The red shading indicates the times between the ionization waves, and the regions shaded in green denote the times when the electric field peaks within the ionization waves.}
\label{fig:efield_drift}
\end{center}
\end{figure}

\begin{figure}
\begin{center}
\includegraphics[width=0.5\columnwidth]{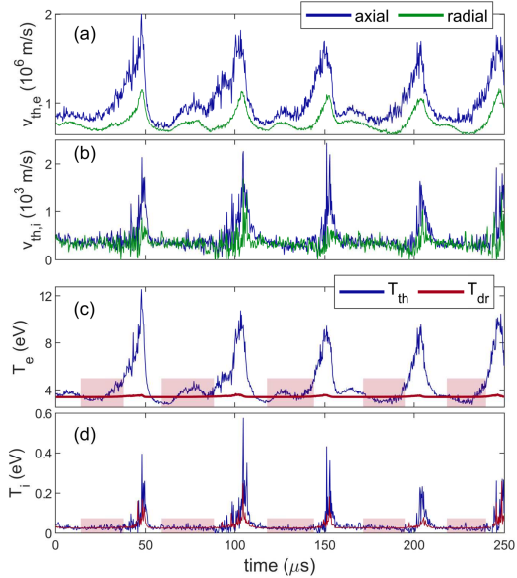}
\caption{(a,b) Thermal velocities for the electrons and ions in the axial (blue) and radial (green) directions calculated from the standard deviation of the velocity distribution. The velocities in the tangential direction (not shown) are similar to those in the radial direction. (c,d) Temperature of the electrons and ions calculated from the thermal velocities, Eq.~(\ref{EqTth}) (blue line), and the average temperature between ionization waves plus the drift energy, Eq.~(\ref{EqTdr}) (red line).  The shaded areas indicate the times between the ionization waves when the ions thermalize with the neutral gas.}
\label{fig:plasma_temp}
\end{center}
\end{figure}

A similar structure is seen in the electron and ion velocities, which rapidly increase in magnitude within an ionization wave.  The velocities are measured from the moments of the velocity distribution.  The first moment is the average velocity, which shows the net mean drift velocity $v_d$ imparted by the DC electric field in the column (figure~\ref{fig:efield_drift}(b,c)).  The second moment of the velocity distribution gives the standard deviation, which is the average (thermal) velocity of the plasma particles, $v_{th}$ (figure~\ref{fig:plasma_temp}(a,b)). The temperatures calculated from the time-dependent thermal velocities, 
\begin{equation}
\label{EqTth}
T_{th} = \frac{2m v_{th}^2}{3 k_B},  
\end{equation} 
are shown in figure \ref{fig:plasma_temp} (c,d). The fully time-averaged electron and ion thermal energies are $\langle\epsilon\rangle_e = 4.4$~eV  and $\langle\epsilon\rangle_i = 0.04$~eV. This is much greater than the average energies calculated between the ionization waves (marked by the shaded regions) $\langle\epsilon\rangle_e = 3.4$~eV and $\langle\epsilon\rangle_i = 0.025~{\rm eV} = 293$~K.  In between the ionization waves, the ions thermalize with the neutral gas at temperature $T_n = 290$~K. 

Note that the drift energy must be carefully taken into account in calculating the mean thermal energy. According to Monte Carlo simulations of ion drift in electric fields, the mean energy of the ions including the drift velocities is given by \citep{RobertsonSternovsky2003}
\begin{equation}
 \frac{1}{2} m_i \langle v_{i}^2 \rangle = \frac{\pi}{4}m_i v_{d,i}^2 + \frac{3}{2} k_B T_n, 
\end{equation}
where $T_n$ is the temperature of the neutral gas, leading to an expression for the ion temperature as a function of the drift velocity \citep{Trottenberg2006}
\begin{equation}
\label{EqTdr}
   T_{dr,i} = T_n + (\frac{\pi-2}{6}) \frac{1}{k_B} m_i v_{dr,i}^2.
\end{equation}

The ion temperature calculated in this manner is shown in figure~\ref{fig:plasma_temp}(d) by the red line.  Applying Eq.~\ref{EqTdr} gives an average ion temperature of $T_{dr,i}$ = 380~K = 0.033~eV over the full time interval, greater than that between the ionization waves, but less than that calculated without the drift correction. Using an equation similar to Eq.~(\ref{EqTdr}) to calculate the average electron energy shows that there is very little difference between the average electron energy between the ionization waves and that averaged over the full time interval, with $T_{dr,e}$ = 39543 K = 3.41 eV (as indicated by the red line in figure~\ref{fig:plasma_temp}(c)).

\clearpage

Since the plasma variations occur on the microsecond timescale and the dust dynamics occur on the millisecond timescale, it seems reasonable to use the time-averaged parameters to set the conditions used in the dust dynamics model. However, the drift velocity plays an important role in determining particle charge and the strength and extent of the ion wake. The ion drift velocity in this case is less than the sound speed of the plasma $c_s = \sqrt{k_v T_e /m_i}$.  Given subsonic ion velocities, as the drift velocity increases the ion current to a grain's surface decreases, causing the grain to become more negatively charged.  An increased charge causes the ions to be more strongly focused, resulting in a stronger ion wake.  The increased flow velocity also causes the spatial extent of the ion wake to be narrower in the radial (cross-stream) direction and extended in the direction of ion flow \citep{Matthews19}. The interparticle interaction energy, as given by Eq. \ref{IvlevPotential}, also depends on the ion flow velocity, predicting that anisotropy in the particle distribution begins to develop when $M_{th} > 0.3$ \citep{Ivlev08, Ivlev11}.  In between the striations, the average ion drift velocity is $v_{d,i}$ = 95~m/s = 0.22~$M_{th}$.  The average drift velocity over all times is 165 m/s = 0.39~$M_{th}$, which would seem to be just great enough to start to induce anisotropy in the particle distribution.  The average value within the ionization waves (times noted by the green boxes in figure \ref{fig:PIC1}), is $\langle v_{d,i} \rangle$ = 489~m/s = 1.14~$M_{th}$, with an average peak value of 1951~m/s = 4.05~$M_{th}$.  Apparently, it during the ionization waves the ion flow is great enough to cause a transition to strongly ordered strings. Accordingly, we are interested in investigating the effect that the increased electric field has on the formation of ordered dust strings in the PK-4 experiment.

\subsection{Dust and ions simulation}

The plasma conditions shown in Figs.~\ref{fig:PIC1} and \ref{fig:efield_drift} are used to model the dynamics and charging of the dust in a flowing plasma using the molecular dynamics code DRIAD (Dynamic Response of Ions and Dust) \citep{Matthews19}, which solves the equations of motion of the ions and dust on their individual time scales. Here we compare the dust dynamics given the time-averaged plasma conditions (Case 1) to three cases where the electron and ion temperatures are set by the temperatures between the ionization waves (denoted by the red shaded regions in Fig.~\ref{fig:efield_drift}), but the electric field is increased to yield different values of the ion drift speed, $M_{th} = v_{dr,i}/v_{th}$. In Case 2, the average axial electric field without the ionization waves present is used (denoted by the red shaded regions in Figs.~\ref{fig:efield_drift} and ~\ref{fig:plasma_temp}). In Case 3, the electric field averaged over the ionization waves (as indicated by the green boxes in Fig.~\ref{fig:efield_drift}) is applied.  In Case 4, the magnitude of the electric field is set by the average of the half-max of the electric field in the ionization waves. In all cases, the polarity switching of the DC electric field is set to 500 Hz with a 50\%  duty cycle (modeling symmetric switching of the electrode polarities) and the average plasma density is set to $n_e = n_i = 2.1 \times 10^{14}$~m$^{-3}$. The electron and ion temperatures, time-varying axial electric field $\tilde{E}$, and resultant time-varying ion drift velocity $\tilde{v}_{dr,i}$ for each case are given in Table~\ref{tab:cases}.

\begin{table}
  \begin{center}
  \def~{\hphantom{0}}
  \begin{tabular}{ ccccc } 
  \hline
  \ Case & 1 & 2 & 3 & 4 \\
  \hline
  $T_e$ (eV,K) & 3.41, 39500  & 3.38, 39200 & 3.38, 39200 & 3.38, 39200 \\ 
  $T_i$ (eV,K) & 0.033, 380  & 0.025, 290 & 0.025, 290 & 0.025, 290 \\ 
  $v_{th,i} (m/s)$ & 489 & 424 & 424 & 424 \\
  $\tilde{E}$ (V/m)  & 245  & 100 & 510 & 1000 \\ 
  $\tilde{v}_{dr,i}$ (m/s)  & 165  & 93 & 429 & 719 \\ 
  $M_{th}$ & 0.34  & 0.22 & 1.01 & 1.69 \\
  \hline
  $\langle Q_d \rangle$ ($e^-$) & 3898 & 3667 & 4191 & 4819 \\
  $\Delta$ ($\mu$m) & 396 & 392 & 401 & 402 \\
  $\langle r \rangle$ ($\mu$m) & 14.5 & 12.6 & 11.6 & 11.9 \\
  $\langle r \rangle / \Delta (\%)$ & 3.6 & 3.2  & 2.9 & 3.0\\
  \hline
  \end{tabular}  
  \caption{Discharge conditions used in the ion and dust simulation and calculated dust charge, inter-particle spacing within the chain, and average radial displacement.}
  \label{tab:cases}
  \end{center}
\end{table}

\section{Dynamics of Ions and Dust}
In each case, we simulate the motion of 20 dust grains (melamine formaldehyde) with radius $a = 3.43~\mu$m, which corresponds to dust particle size available in the PK-4 experiment. The dust particles are initially placed in a cloud near the center of the simulation region, which has a radius of 1.5~$\lambda_e$ and length of 12~$\lambda_e$, where $\lambda_e = 940~\mu$m is the electron Debye length of the plasma calculated for Cases 2-4.  The equation of motion for the dust grains with mass $m_d$ and charge $Q_d$  is given by 
\begin{equation}
m_d  \frac{d \vec {v}_d}{dt}= \vec{F}_{dd} + \vec{F}_{id} + Q_d \tilde {E} + \nu^2 Q_d r\hat{r} - \beta \vec {v} + \zeta(t).
\label{eqn_dust}
\end{equation}

The forces between the dust particles $\vec{F}_{dd}$ are Coulomb interactions, as the ions in the simulation provide the shielding, while the forces between the dust and ions $\vec{F}_{id}$ are taken to be Yukawa interactions \citep{Matthews19, Ashrafi20}. The ion-dust interactions are accumulated over the elapsed ion timesteps and then averaged before calculating the dust acceleration. The electric field $\tilde{E}$ is the axial electric field in the DC plasma which switches direction with the polarity switching frequency. There is a very strong confining force to keep the particles from the ends of the simulation region where the ions are injected (the ions need to travel approximately one Debye length to reach their equilibrium distribution). The parabolic radial confinement potential approximates the electric field from surrounding chains of charged dust particles where the confining strength $\nu^2 \propto \bar{Q}/(4 \pi \epsilon_0 \Delta^3)$, $\bar{Q}$ and $\Delta$ are the average expected particle charge and particle separation, and a constant of proportionality is used to account for the fact that there are multiple chains providing the confinement. Depending on the number of nearest neighbors assumed to participate in the confinement and the shielding length of the interaction potential, this constant of proportionality can range from $C = 0.5 - 4.5$.   Dust density wave experiments performed in the PK-4 in neon gas at 40 Pa found an estimated particle charge of $Z_{d} \approx 2200$ for $a$ = 1.60 $\mu$m particles \citep{Jaiswal18}; assuming the charge scales linearly with the dust radius, the charge on a particle with radius $a = 3.43~\mu$m is estimated to be $Z_{d} \approx $ 4500. The average interparticle spacing, estimated from the number of particles visible in an image frame from the PK-4 experiment, is $\Delta \approx$ 305 $\mu$m. In all four cases simulated here, a fixed value of $\nu^2 = 3.0 \bar{Q}/(4 \pi \epsilon_0 \Delta^3) = 6.8 \times 10^{5}$~Vm$^{-2}$ was used. The neutral gas  (density $n_g$ and molecular mass $m_g$) provides both an energy sink and source with the neutral gas drag depending on the drag coefficient $\beta$ = $ (4\pi/3) \delta a^2 n_g m_g \sqrt{8k_B T_g / \pi m_g}$ (where $\delta$ is a material-dependent constant in the range of 1.0 - 1.44; here we used 1.44 to represent diffuse reflection with accommodation of gas molecules from a non-conductor) and a Langevin thermostat set by $\zeta = \sqrt{2\beta k_B T_g / \Delta t_d}$  (the dust time step $\Delta t_d$ = 0.1~ms). The system is allowed to evolve for 1.8~s, at which time the dust particles have reached their equilibrium configuration.  

The wakefield interactions are included self-consistently by solving the equations of motion for the ions 

\begin{equation}
m_d  \frac{d \vec {v}_i}{dt}= q_i \vec {E} + \vec{F}_{ii} +\vec{F}_{id},
    \label{eqn_ion}
\end{equation}
where the electric field consists of the confining electric field found within a cylindrical cavity within a homogeneous distribution of background ions, as well as the electric field in the DC plasma with polarity switching, $\tilde{E}$.  The ion-ion interactions  $\vec{F}_{ii}$ are derived from a Yukawa potential where the shielding is provided by the electrons, whereas the force between the ions and dust $\vec{F}_{id}$ is taken to be Coulombic in nature. This asymmetric treatment of the dust-ion forces has been shown to give a reasonable quantitative agreement for the potential distribution and interparticle forces \citep{Piel17}. The ions reach equilibrium on a time scale comparable to the ion plasma period $2\pi/\omega_{i} = 2\pi/\sqrt{n_i e^2/\epsilon_0 m_i} = 3.0 \times 10^{-6}$~s, which is fast compared to the period of the polarity switching, 2 ms.  The effect of ion-neutral collisions are incorporated using the null collision method \citep{Donko11}.

The charge on the dust grains is calculated self-consistently within the plasma wakefield by summing the ion collisions over the elapsed ion timesteps to determine the ion current. The electrons are assumed to have a Boltzmann distribution and the electron current is set using orbital-motion-limited (OML) theory.

\section{Results}

\begin{figure}
\begin{center}
\includegraphics[width=0.5\columnwidth]{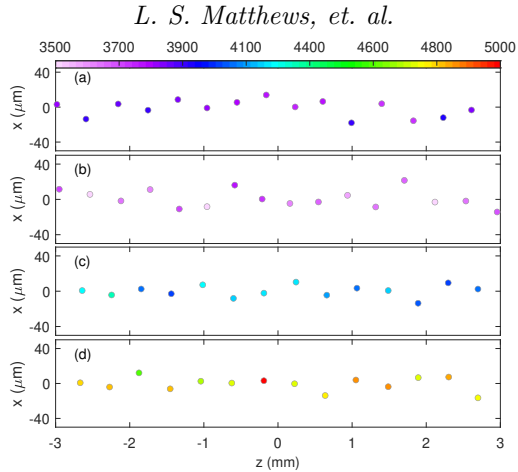}
\caption{Final equilibrium charge and dust configuration for each case. Note the scale in x is magnified to show the displacement from the central axis. The color bar indicates the charge in units of elementary charge $e^-$. (a) Case 1 $\langle Q_d \rangle$ = 3900  $e^-$, (b) Case 2 $\langle Q_d \rangle$ = 3670  $e^-$, (c) Case 3 $\langle Q_d \rangle$ = 4190  $e^-$, (d) Case 4 $\langle Q_d \rangle$ = 4820  $e^-$. }
\label{dust_charge}
\end{center}
\end{figure}

The resulting equilibrium dust charge and spatial configuration of the dust are shown  for the four cases in figure~\ref{dust_charge}. The view shown is a projection into the xz-plane, with the radial scale magnified to show the relative displacement from the central axis.  The ion-gas collisions cause the negative charge of the particles in the chains (indicated by the marker color) to be reduced from that predicted by OML theory, but the negative charge state increases with the ion drift speed, as expected.    

\begin{figure}
\begin{center}
\includegraphics[width=0.5\columnwidth]{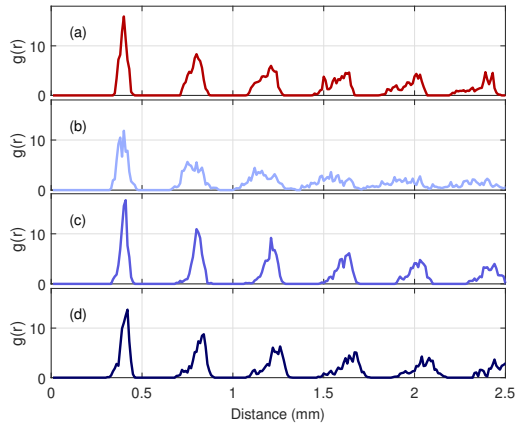}
\caption{Pair correlation functions averaged over 0.5 s for each case.  (a) Case 1,  (b) Case 2, (c) Case 3, (d) Case 4. }
\label{g_r}
\end{center}
\end{figure}

The degree of order in the chain is evaluated using the linear pair correlation function $g(r)$, which was calculated at each dust time step and then averaged over the last 5000 time steps (0.5 s). The results are shown in figure~\ref{g_r}. In general, the order within the chain increases as the electric field is increased (Cases 2-4). Case 2 (figure~\ref{g_r}b) shows very little order beyond the third peak. This is a clear indication that the enhanced wakes due to the strong ion flow in the ionization waves contribute to the formation of ordered chains.  The fully time-averaged condition, Case 1, figure~\ref{g_r}(a), leads to a configuration which is more ordered than the thermal plasma without ionization waves (Case 2), but less ordered than the other two cases.  Interestingly, the case with the highest degree of order is not that with the greatest electric field and resultant ion flow, but Case 3, figure~\ref{g_r}(c), which employs the electric field averaged over the ionization waves. 

\begin{figure}
\begin{center}
\includegraphics[width=0.5\columnwidth]{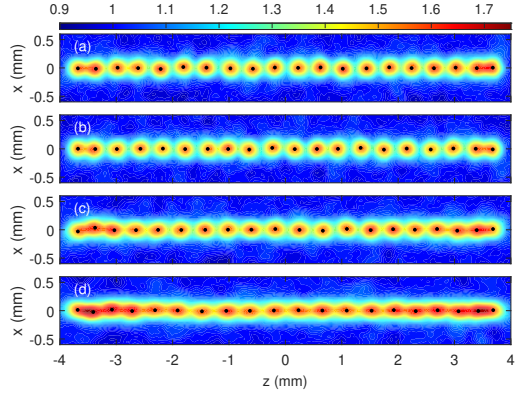}
\caption{Final equilibrium ion density, where the colorbar indicates the density in units of the background density $n_i$/$n_{i0}$. The ion densities are shown for a slice through the xz-plane averaged over 20 polarity cycles (40 ms). (a) Case 1, (b) Case 2, (c) Case 3, (d) Case 4. }
\label{ion_dens}
\end{center}
\end{figure}

\begin{figure}
\begin{center}
\includegraphics[width=0.5\columnwidth]{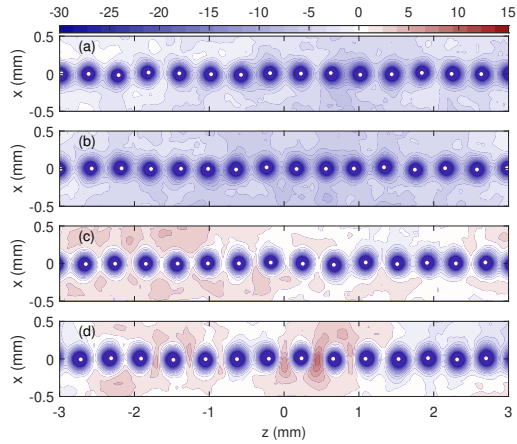}
\caption{Equilibrium electric potential, where the colorbar indicates the difference from the maximum positive potential upstream/downstream of the strings in mV. The potentials are shown for a slice through the xz-plane averaged over 20 polarity cycles  (40~ms). (a) Case 1, (b) Case 2, (c) Case 3, (d) Case 4. }
\label{potential}
\end{center}
\end{figure}

Some clues to this order can be found by examining the ion density at equilibrium and the overall electric potential of the system.  In figure~\ref{ion_dens} and figure~\ref{potential}, the ion density and the electric potential are shown for a slice in the xz-plane.  As shown in figure~\ref{ion_dens}, each dust particle attracts a cloud of ions.  In Cases 1 and 2 with the averaged plasma conditions, a distinct ion cloud surrounds each particle.  As the electric field is increased in Cases 3-4, the ion cloud becomes elongated in the axial direction and the cloud around a grain begins to merge with that of neighboring grains.  The increased particle charge, in addition to the increased ion flow speed, concentrates the ions in a high-density ridge along the dust string.  

\begin{figure}
\begin{center}
\includegraphics[width=0.5\columnwidth]{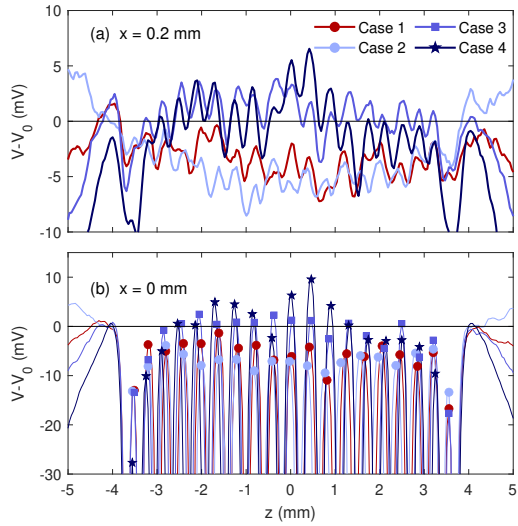}
\caption{Equilibrium electric potential along the axial direction averaged over 20 polarity cycles  (40~ms). In (a), the profile is shown just outside the dust string along $x = 0.2$~mm, in (b) the profile is shown along the center of the dust string at $x = 0.0$~mm. The maximum positive potential between the dust grains are marked with symbols for each of the four cases.}
\label{potential_profile}
\end{center}
\end{figure}

Finally, the combined electric potential from the ions and dust is shown at equilibrium in figure~\ref{potential}.  Note that these figures are zoomed in on the central portion of the chain.  The potentials are averaged over 20 polarity cycles (0.04~s). The potential is measured with respect to the maximum potential just upstream/downstream of the dust string at $z \approx \pm$ 4~mm.  Profiles of the total potential along the axial direction are compared in figure~\ref{potential_profile} just above the dust string (in the radial direction) at $x = 0.2$~mm (figure~\ref{potential_profile}a) and along the center of the dust chain at $x = 0.0$~mm (figure~\ref{potential_profile}b). Note that in Cases 1 and 2 the overall potential is dominated by the dust grains and is negative over much of the region surrounding the string.  In Case 3, the potential is slightly positive just to the outside of the dust chain.  In Case 4, an alternating positive/negative potential structure starts to emerge along the length of the chain. 

\clearpage

\section{Discussion}
\begin{figure}
\begin{center}
\includegraphics[width=0.5\columnwidth]{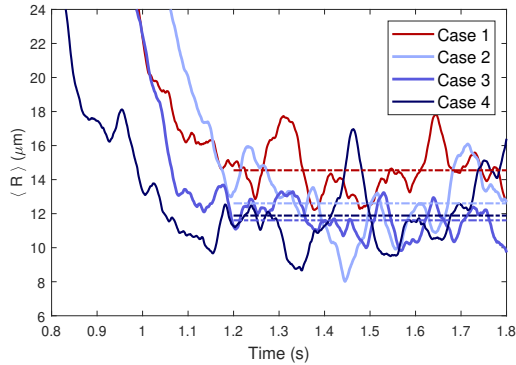}
\caption{Average radial displacement of particles in the numerical simulation. The dashed lines indicate the average radial displacement over the last 0.6 s, after all of the particles have reached their equilibrium configuration. At equilibrium, the average radial displacements are 14.5 $\mu$m, 12.6 $\mu$m, 11.6 $\mu$m, and 11.9 $\mu$m for cases 1-4, respectively. }
\label{radial_position}
\end{center}
\end{figure}

 It is expected that the radial confinement should be proportional to $Q_d^2$, assuming that the radial confinement is due to the interaction between neighboring strings. As given in Eq.~\ref{eqn_dust}, the magnitude of the radial confining force is $\nu^2 Q_d r$.  For simplicity, the constant $\nu^2$  was set to be $6.8 \times 10^5$~Vm$^{-2}$ for all of the cases, resulting in a restoring force which is proportional to $Q_d$. Thus the radial restoring force used can be considered to be underpredicted for Case 3 and 4 and overpredicted for  Case 2, relative to Case 1. The average radial position of all the particles in each chain as a function of time is shown in figure~\ref{radial_position}. After the initial Coulomb expansion of the dust cloud at the beginning of the simulation, the particles all settle near the z-axis.  As expected, the case with the largest average particle charge (Case 4) experiences the greatest radial restoring force and reaches the equilibrium radial position most quickly, followed by the other cases in order of decreasing average particle charge. However, even though the particles in Cases 3 and 4 have the greatest average charge, these chains have the smallest average radial displacement, representing better string structure.  Notably, Case 1, with the time-averaged plasma conditions, has the greatest average radial displacement.  This is a clear indication that the ion focusing produced by the strong axial electric field in Cases 3 and 4 allows for a smaller inter-particle spacing within the string, despite the increased particle charge, and enhances string alignment.

\begin{figure}
\begin{center}
\includegraphics[width=0.5\columnwidth]{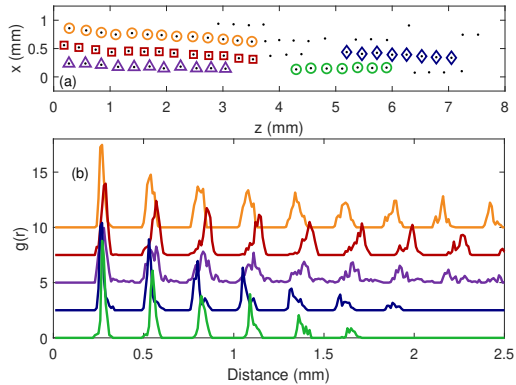}
\caption{(a) Chains of dust particles observed in the PK-4 experiment. Symbols mark particles in five different chains which remained intact over the full time period. (b) Linear pair correlation function for each chain marked in (a), averaged over 70 frames (1.0~s).  Each line is successively offset by 2.5 for clarity.}
\label{PK4_data}
\end{center}
\end{figure}

For comparison, data from Campaign 4 of the PK4 experiment performed in February, 2017, is shown in figure~\ref{PK4_data}a showing chains consisting of 6.86~$\mu$m-diameter particles which were observed in neon gas at 45 Pa. The particles were trapped by the polarity-switched discharge (current 1.0 mA and frequency 500 Hz) with a duty cycle of 0.72, in order to compensate for the residual gas flow. A duty cycle of 0.72 corresponds to an asymmetric AC mode with 72\% of the cycle at positive voltage and 28\% at negative voltage.  The linear pair correlation function for five different chains (marked by the different symbols), were calculated and averaged over 70 frames (1.0 s) as shown in figure \ref{PK4_data}b. Qualitatively, the pair correlation functions most closely resemble that shown for Case 3 (figure \ref{g_r}c) in that there are distinct, separate peaks out to the position of the sixth-nearest neighbor.  The average inter-particle spacing for the five chains is $\Delta = 270, 282, 281, 270, 277~\mu$m, from top to bottom, respectively, calculated from the first peak in g(r). The average radial displacements of a chain's particles, measured as the perpendicular distance from a linear fit to the positions of the particles in a chain are $\langle r \rangle = 11.6, 20.4, 16.8, 13.2, 17.8 ~\mu$m. In the experiment, the average inter-particle spacing within the chain is smaller, and average radial displacement is larger, than that found for the four cases in the numerical model (see Table~\ref{tab:cases}), such that $\langle r \rangle / \Delta$ = 4.3, 7.2, 6.0, 4.9, 6.4\% for each chain, respectively.  This suggests that the particle charge in the experiment may be less than estimated, possibly due to the fact that the dust density is great enough to deplete the electrons in the vicinity of the dust cloud.  This would result in stronger ion wake potentials along the chain axis and weaker repulsion between neighboring chains, allowing the particles more freedom for radial displacements.  Another possible explanation for the observed larger average radial displacement of the dust particles in the experiment, as compared to the simulation, is that the asymmetric duty cycle used in the experiment lead to asymmetric ion focusing around the dust grains.  This could produce a stronger positive wake on one side of the dust, allowing smaller intra-chain particle spacing, while weakening the radial restoring force and resulting in less stable chains. 

\section{Conclusion}

A simulation of dust dynamics within a DC discharge plasma was used to investigate the role of strong electric fields created by ionization waves on the formation of chain-like dust structures within the plasma.  A PIC/MCC code was used to determine the plasma conditions within the discharge tube, which were used to set the initial conditions and boundary conditions for an N-body simulation resolving the motion of the ions and the dust. The PIC/MCC simulation revealed that there are very strong variations in the plasma conditions on the microsecond scale which result in a large axial electric field with a peak magnitude of about 2000~V/m, about 20 times greater than the background value between the ionization waves.  Simulations of dust charging and dynamics show that time-averaged plasma temperatures and axial electric field lead to a weakly ordered string structure, leading to the conclusion that the time-averaged conditions don't seem to fully capture the plasma conditions which lead to chain formation. However, simulations using the plasma temperatures and densities between the ionization waves with an applied axial electric field show that the order within the string increases with the electric field strength.  The numerical results most closely resemble data from the PK-4 experiment when the average electric field during the ionization waves is applied.  It appears that the enhanced electric field associated with the ionization waves could play an important role in generating the string-like structures observed in the PK-4 experiment. 

 These simulations were run assuming constant plasma conditions including electron and ion temperatures and number densities. Future work will examine the effect of the time-varying plasma parameters calculated from the PIC/MCC simulation on the dust charging and dynamics.

All authors gratefully acknowledge the joint ESA -- Roscosmos ``Experiment Plasmakristall-4'' on-board the International Space Station. The microgravity research is funded by the space administration of the Deutsches Zentrum f\"ur Luft- und Raumfahrt e.V. with funds from the federal ministry for economy and technology according to a resolution of the Deutscher Bundestag under Grants No. 50WM1441 and No. 50WM2044. A. M. Lipaev and A. D. Usachev were supported by the Russian Science Foundation Grant No. 20-12-00365 and participated in preparation of this experiment and its execution on board the ISS. L. S. Matthews, T.W. Hyde and M. Rosenberg received support from NASA Grant number 1571701 and NSF Grant numbers 1740203 (LSM, TWH, and MR) and the US Department of Energy, Office of Science, Office of Fusion Energy Sciences under award number DE-SC-0021334 (LSM and TWH).  P. Hartmann  gratefully acknowledges support from the Hungarian Research, Development and Innovation Office via grant K-134462. 

\bibliographystyle{jpp}

\bibliography{neonbib}

\end{document}